\documentclass[]{JHEP3}

\usepackage{graphicx}
\usepackage{amsmath}

\title{More on QCD Ghost Dark Energy}

\author{Rong-Gen Cai, Zhong-Liang Tuo\\
    State Key Laboratory of Theoretical Physics,
    Institute of Theoretical Physics, Chinese Academy of Sciences,
    P.O. Box 2735, Beijing 100190, China\\
    E-mail:
    \email{cairg@itp.ac.cn,tuozhl@itp.ac.cn}}
\author{Ya-Bo Wu, Yue-Yue Zhao\\
    Department of Physics, Liaoning Normal University, Dalian, 116029,
    China\\
    E-mail:
    \email{ybwu61@163.com,zhaoyueyue198737@163.com}}

\received{\today}       
\accepted{\today}       

\abstract{
The difference between vacuum energy of quantum fields in
Minkowski space and in Friedmann-Robterson-Walker universe might be
related to the observed dark energy. The vacuum energy of the
Veneziano ghost field introduced to solve the $U(1)_A$ problem in
QCD is of the form, $ H+ {\cal O}(H^2)$. Based on this, we study the
dynamical evolution of a phenomenological dark energy model whose
energy density is of the form $\alpha H+\beta H^2$. In this model,
the universe approaches to a de Sitter phase at late times. We fit
the model with current observational data including SnIa, BAO, CMB,
BBN, Hubble parameter and growth rate of matter perturbation. It
shows that the universe begins to accelerate at redshift $z\sim
0.75$ and this model is consistent with current data. In particular,
this model fits the data of growth factor well as the $\Lambda CDM$
model.
}

\begin{document}
\maketitle

\section{Introduction}

The accelerating expansion is still a mystery of modern cosmology
since its discovery in 1998~\cite{Riess:1998cb}, and a new energy
component called dark energy (DE) is needed to explain this
acceleration expansion within the framework of general relativity.
The simplest model of DE is the cosmological constant, which is
consistent with all observational data, but it faces with the fine
tuning problem~\cite{finetunning}. Instead, many alternative DE
models have also been proposed
\cite{Copeland:2006wr,Caldwell:1997mh,Steinhardt:1999nw,Capozziello:2003tk,Li:2004rb,Nojiri:2005sr,Nojiri:2003ft},
but almost all of them explain the acceleration expansion either by
introducing new degree(s) of freedom or by modifying general
relativity.

Recently the so-called QCD ghost dark energy has been proposed
in~\cite{Urban,Ohta:2010in,CTZS}. The Veneziano ghost field plays a
crucial role in the resolution of the  $U(1)_{A}$ problem in QCD
~\cite{QCDghost}. The ghost field has no contribution to the vacuum
energy density in Minkowski spacetime, but in a curved spacetime, it
gives rise to a  vacuum energy density proportional to
$\Lambda_{QCD}^{3}H$~\cite{Zhitnitsky:2010ji,Holdom:2010ak,Zhitnitsky:2011tr,Thomas:2011ee},
where $\Lambda_{QCD}$ is QCD mass scale and $H$ is Hubble parameter.
Note that in this ghost dark energy model, there are no unwanted
features such as violation of gauge invariance, unitarity, causality
etc.~\cite{Urban,Ohta:2010in,Zhitnitsky:2010ji,Holdom:2010ak,Zhitnitsky:2011tr,Thomas:2011ee}.
In fact, the description in terms of the Veneziano ghost is just a
matter of convenience to describe very complicated infrared dynamics
of strongly coupled QCD. The Veneziano ghost is not a physical
propagating degree of freedom, one can describe the same dynamics
using some other approaches (e.g. direct lattice simulations)
without using the ghost. Therefore the Veneziano ghost field is
quite different from those ghost fields in some dark energy models
in the literature, there those ghost fields are real physical
degrees of freedom, introduced in order to have the equation of
state of dark energy to cross $-1$. On the other hand, the vacuum
energy is generally expected exponentially suppressed because QCD is
a theory with a mass gap. However, this issue is elaborated in some
details in
\cite{Zhitnitsky:2010ji,Holdom:2010ak,Zhitnitsky:2011tr,Thomas:2011ee},
there it has been convincingly argued that the complicated
topological structure of strongly coupled QCD will lead to a vacuum
energy density with an inverse linear scale of manifold size in a
nontrivial background. The power law behavior is also supported by
recent lattice result \cite{Holdom:2010ak,lattice}. Very recently it
has been shown that this behavior is also got supported from the
holographic description of gauge field~\cite{Zhit}.

Because this model is totally embedded in standard model and general
relativity, one needs not to introduce any new degrees of freedom or
to modify Einstein's general relativity. In this model, the energy
density of DE is roughly of order $\Lambda_{QCD}^{3}H$, with
$\Lambda_{QCD}\sim100MeV$ and $H\sim10^{-33}eV$, so
$\Lambda_{QCD}^{3}H$ gives the right order of observed DE energy
density. This numerical coincidence is impressive and also means
that this model gets rid of fine tuning problem
\cite{Urban,Ohta:2010in}. The model parameters have been fitted
recently by observational data including SnIa, BAO, CMB, BBN and
Hubble parameter data~\cite{CTZS}. It shows that this model is
consistent with those observational data.

On the other hand, it is convincingly argued that the contribution
of zero-point fluctuations of quantum field to the total energy
density should be computed by subtracting the Minkowski space result
from the computation in a FRW
space-time~\cite{Magg,Maggiore:2011hw}. Usually the difference, $H^2
\Lambda_c^2$, between the vacuum energies in Minkowski space and in
the FRW space-time, is absorbed into a renormalization of Newton's
constant $G$, here $H$ is the Hubble constant of the FRW universe
and $\Lambda_c$ is the cutoff. However, it is true only under the
assumption that the vacuum expectation value of the energy-momentum
tensor is conserved in isolation~\cite{Maggiore:2011hw}. In that
reference the authors investigated the role of this term as early
dark energy in the evolution of the universe.

Notice the fact that the vacuum energy from the Veneziano ghost
field in QCD is of the form $H + {\cal O}(H^2)$, see for example,
\cite{Zhit}, while in the previous works on the QCD ghost dark
energy model, the leading term $H$ has been considered only. Having
considering the study in \cite{Maggiore:2011hw}, one may expect that
the subleading term $H^2$ in the ghost dark energy model might play
a crucial role in the early evolution of the universe, acting as the
early dark energy.

 Based on the QCD ghost dark energy, in this work  we therefore investigate a
phenomenological model with energy density $\rho_{DE} = \alpha H+
\beta H^2$. For other motivations to consider this form
see~\cite{varying}, there it is argued that this form of varying
cosmological constant could be a possible candidate to solve the two
fundamental cosmological puzzles. We study the cosmological
evolution of this DE model, fit this model with current
observational data and give constraints on the model parameters.
Besides, it is worth noticing that a varying DE should have some
effect on the evolution of the matter perturbation, so we study the
first order perturbation to the matter density and fit this model
with the data of linear growth factor.

The paper is organized as follows. In Section 2 we study the
dynamical evolution of the DE model. In Section 3, we fit this model
with current observational data and discuss the fitting results. The
data used are Union II SnIa sample \cite{Amanullah:2010vv}, BAO data
from SDSS DR7 \cite{Percival:2009xn}, CMB data $(R,\, l_{a},\,
z_{*})$ from WMAP7 \cite{Komatsu:2010fb}, 12 Hubble evolution data
\cite{Simon:2004tf,Gaztanaga:2008xz} and Big Bang Nucleosynthesis
(BBN) \cite{Serra:2009yp,Burles:2000zk}, and we also study the
effect of the DE on the linear perturbation of matter density. We
summarize our work and give some discussions in Section 4.


\section{Dynamics of QCD Ghost Dark Energy}

To study the dynamics of the DE model, we consider a flat FRW
universe with three energy components: matter, DE and radiation. In
this ghost DE model, the energy density of DE is given by
$\rho_{DE}=\alpha H+\beta H^2$, where $\alpha$ is a constant with
dimension $[energy]^{3}$, roughly of order of $\Lambda_{QCD}^{3}$
where $\Lambda_{QCD}\sim100\text{MeV}$ is QCD mass scale, and
$\beta$ is another constant with dimension $[energy]^{2}$. For
convenience, we define $\gamma\equiv 1-\frac{8 \pi G}{3}\beta$ and
use this throughout the paper.

Arming with this DE density, the Friedman equation
reads\begin{equation} H^{2}=\frac{8\pi G}{3\gamma}\left(\alpha
H+\rho_{m}+\rho_{r}\right),\label{eq:friedman}\end{equation}
 where $\rho_{m}$ is energy density of matter, whose continuity
 equation gives \begin{equation}
\dot{\rho}_{m}+3H\rho_{m}=0\;\Longrightarrow\;\rho_{m}=\rho_{m0}a^{-3}.
\label{eq:matter}\end{equation} and $\rho_{r}$ is energy density of
the radiation, whose continuity
 equation gives \begin{equation}
\dot{\rho}_{r}+4H\rho_{r}=0\;\Longrightarrow\;\rho_{r}=\rho_{r0}a^{-4}.
\label{eq:rad}\end{equation}
 We have set $a_{0}=1$ and the subscript $0$ stands for the present value of some quantities.
Solving the Friedman equation, we have
  \begin{equation}
H_{\pm}=\frac{4\pi G}{3\gamma}\alpha\pm\sqrt{\left(\frac{4\pi
G}{3\gamma}\alpha\right)^{2} +\frac{8\pi
G}{3\gamma}\rho_{m0}a^{-3}+\frac{8\pi G}{3\gamma}\rho_{r0}a^{-4}}.
\label{eq:hubble}\end{equation} There are two branches, $H_{+}$
represents an expansion solution, while $H_{-}$  a contraction one.
We neglect the latter since it goes against the observation, and for
simplicity, write $H_{+}$ as $H$ in what follows.

Expressed with fraction energy density of matter and radiation,
$\Omega_{m0},\,\Omega_{r0}$, Equation \ref{eq:hubble} gives an
important constraint among these parameters:
\begin{equation}
(\gamma -\Omega_{m0}-\Omega_{r0})H_0=\frac{8\pi G}{3}\alpha.
\label{eq:relation}\end{equation}
 Further, Equation \ref{eq:hubble} can be rewritten as
\begin{equation}
H(z)=H_0\left (
\kappa+\sqrt{\kappa^2+\frac{\Omega_{m0}(1+z)^3+\Omega_{r0}(1+z)^4}{\gamma}}\right
). \label{eq:hz}\end{equation}
 where $\kappa = (1-(\Omega_{m0}+\Omega_{\gamma0})/\gamma)/2$ and  $z$ is the redshift,
$z=1/a-1$. We can see from (\ref{eq:hz}) that the universe
approaches to a de Sitter phase with Hubble parameter $2\kappa H_0$
at late times, while it is dominated by matter and radiation terms
at early times.

Note that there exist many papers focusing on the coupling between
the time-dependent vacuum energy and matter~\cite{couple}. But we do
not consider here such coupling in our analysis. Namely in our
discussion, the DE, matter and radiation are separately conserved.
In that case, the corresponding time evolution equation for the
matter density contrast $D\equiv \delta\rho_m/\rho_m$ is given by:
\begin{equation}
\ddot{D}+2H\dot{D}-4\pi G\rho_mD=0, \label{contrast}\end{equation}
where the over dot denotes the derivative with respect to the cosmic
time. In terms of the growth factor~\cite{growth-factor}, Equation
\ref{contrast} can be rewritten as
\begin{equation}
-(1+z)H(z)^2\frac{df}{dz}+2H(z)^2f+H(z)^2f^2-(1+z)H(z)\frac{dH(z)}{dz}f=\frac{3\Omega_{m0}(1+z)^3}{2},
\label{growth}\end{equation}
 where the growth factor $f$ is defined as
$f=-(1+z)\frac{d\ln D}{dz}$. In general, there is no analytical
solution to Equation \ref{growth}, and we need to solve it
numerically. But it is very interesting that the solution of the
equation can be approximated as~\cite{approximate}
\begin{equation}
f=\Omega_{m}(z)^\lambda,
\end{equation}
and the growth index $\lambda$ can be obtained for some general
models as
\begin{equation}
\lambda=\frac{3}{5-\frac{w}{1-w}}+\frac{3}{125}\frac{(1-w)(1-3w/2)}{(1-6w/5)^3}(1-\Omega_{m}(z)).
\end{equation}
where $w$ is the equation of state of DE. For the case with
$1-\Omega_m(z)$ being between zero and $0.8$, the accuracy is better
than $1\%$. For the $\Lambda CDM$ model, the approximation
$f(z=0)=\Omega_{m0}^{0.6}+\Omega_{\Lambda 0}(1+\Omega_{m0}/2)/70$
can be made~\cite{Lahav}. But in our analysis, instead of
parametrization of $\lambda$~\cite{parametrization}, we will solve
the Equation \ref{growth} numerically, by setting the initial
condition $f(z=0)=f_0$, where $f_0$ is a free parameter to be
constrained by observational data.

\section{Data Fitting}

\subsection{Model}

In order to fit the model with current observational data, we
consider a more realistic model which includes DE, Cold Dark Matter,
radiation and baryon in a flat FRW universe in this section. In this
case, the dimensionless Hubble parameter can be written as,
\begin{eqnarray} E & \equiv &
\frac{H}{H_{0}}=\kappa
+\sqrt{\kappa^2+\frac{\Omega_{m0}(1+z)^3+\Omega_{r0}(1+z)^4}{\gamma}},\end{eqnarray}
where the energy density of baryon and Cold Dark Matter are always
written together as $\Omega_{DM0}+\Omega_{b0}=\Omega_{m0}$, and
$\Omega_{DM0},\Omega_{b0},\Omega_{r0}$ are present values of
dimensionless energy density for Cold Dark Matter, baryon and
radiation, respectively.  The energy density of radiation is the sum
of those of photons and relativistic neutrinos
\[ \Omega_{r0}=\Omega_{\gamma0}\left(1+0.2271N_{n}\right),\]
 where $N_{n}=3.04$ is the effective number of neutrino species and $\Omega_{\gamma0}=2.469\times10^{-5}h^{-2}$
for $T_{cmb}=2.725K$ ($h={H_{0}}/{100}\, Mpc\cdot km\cdot s^{-1}$).

We will choose $h,\, \gamma,\, \Omega_{b0}$ and $\Omega_{m0}$ (and
also $f_0$ when we consider the growth factor) as free parameters of
the model in the following data fitting. This relation Equation
\ref{eq:relation} implies that there exists a strong degeneracy
among  $h,\, \gamma$ and $\Omega_{m0}$.

\subsection{Observational Datasets}

We fit our model by employing some observational data including
SnIa, BAO, CMB, Hubble evolution data, BBN and the data of growth
factor.

The data for SnIa are the 557 Uion II sample \cite{Amanullah:2010vv}.
$\chi_{sn}^{2}$ for SnIa is obtained by comparing theoretical distance
modulus $\mu_{th}(z)=5\log_{10}[(1+z)\int_{0}^{z}dx/E(x)]+\mu_{0}$
($\mu_{0}=42.384-5\log_{10}h$) with observed $\mu_{ob}$ of supernovae:
\begin{equation}
\chi_{sn}^{2}=\sum_{i}^{557}\frac{[\mu_{th}(z_{i})-\mu_{ob}(z_{i})]^{2}}{\sigma^{2}(z_{i})}.
\end{equation}
 To reduce the effect of $\mu_{0}$, we expand $\chi_{sn}^{2}$ with
respect to $\mu_{0}$ \cite{Nesseris:2005ur} : \begin{equation}
\chi_{sn}^{2}=A+2B\mu_{0}+C\mu_{0}^{2}\label{eq:expand}\end{equation}
 where \begin{eqnarray*}
A & = & \sum_{i}\frac{[\mu_{th}(z_{i};\mu_{0}=0)-\mu_{ob}(z_{i})]^{2}}{\sigma^{2}(z_{i})},\\
B & = & \sum_{i}\frac{\mu_{th}(z_{i};\mu_{0}=0)-\mu_{ob}(z_{i})}{\sigma^{2}(z_{i})},\\
C & = & \sum_{i}\frac{1}{\sigma^{2}(z_{i})}.\end{eqnarray*}
 (\ref{eq:expand}) has a minimum as
 \begin{equation}
\widetilde{\chi}_{sn}^{2}=\chi_{sn,min}^{2}=A-B^{2}/C,
\end{equation}
 which is independent of $\mu_{0}$. In fact, it is equivalent to
performing an uniform marginalization over $\mu_{0}$, the difference
between $\widetilde{\chi}_{sn}^{2}$ and the marginalized $\chi_{sn}^{2}$
is just a constant \cite{Nesseris:2005ur}. We will adopt $\widetilde{\chi}_{sn}^{2}$
as the goodness of fit between theoretical model and SnIa data.

The second set of data is the Baryon Acoustic Oscillations (BAO)
data from SDSS DR7 \cite{Percival:2009xn}, the datapoints we use are
\[
d_{0.2}=\frac{r_{s}(z_{d})}{D_{V}(0.2)}\]
 and
\[
d_{0.35}=\frac{r_{s}(z_{d})}{D_{V}(0.35)},\]
 where $r_{s}(z_{d})$ is the comoving sound horizon at the baryon
drag epoch \cite{Eisenstein:1997ik}, and

\[
D_{V}(z)=\left[\left(\int_{0}^{z}\frac{dx}{H(x)}\right)^{2}\frac{z}{H(z)}\right]^{1/3}\]
 encodes the visual distortion of a spherical object due to the non
Euclidianity of a FRW spacetime. The inverse covariance matrix of
BAO is
\begin{eqnarray*} C_{M,bao}^{-1} & = & \left(\begin{array}{ccc}
30124 & -17227 \\
-17227 & 86977\end{array}\right).\end{eqnarray*}
The $\chi^2$ of the
BAO data is constructed as:
\begin{equation}
\chi_{bao}^{2}=Y^{T}C_{M,bao}^{-1}Y,
 \end{equation}
  where
\[
Y=\left(\begin{array}{c}
d_{0.2}-0.1905\\
d_{0.35}-0.1097\end{array}\right).\]

The third set of data we use are CMB datapoints ($R,l_{a},z_{*}$)
from WMAP7 \cite{Komatsu:2010fb}. $z_{*}$ is the redshift of
recombination \cite{Hu:1995en}, $R$ is the scaled distance to
recombination

\[
R=\sqrt{\Omega_{m0}}\int_{0}^{z_{*}}\frac{dz}{E(z)},\]
 and $l_{a}$ is the angular scale of the sound horizon at recombination

\[
l_{a}=\pi\frac{r(a_{*})}{r_{s}(a_{*})},\]
 where $r(z)=\int_{0}^{z}dx/H(x)$ is the comoving distance and $r_{s}(a_{*})$
is the comoving sound horizon at recombination \[
r_{s}(a_{*})=\int_{0}^{a_{*}}\frac{c_{s}(a)}{a^{2}H(a)}da,\]
 where the sound speed $c_{s}(a)=1/\sqrt{3(1+\overline{R}_{b}a)}$
and $\overline{R}_{b}=3\Omega_{b}^{(0)}/4\Omega_{\gamma}^{(0)}$ is
the photon-baryon energy density ratio.
The $\chi^{2}$ of the CMB
data is constructed as:
 \begin{equation}
\chi_{cmb}^{2}=X^{T}C_{M,cmb}^{-1}X,
\end{equation}
 where
\[
X=\left(\begin{array}{c}
l_{a}-302.09\\
R-1.725\\
z_{*}-1091.3\end{array}\right)\] and the inverse covariance matrix
\begin{eqnarray*} C_{M,cmb}^{-1} & = & \left(\begin{array}{ccc}
2.305 & 29.698 & -1.333\\
29.698 & 6825.270 & -113.180\\
-1.333 & -113.180 & 3.414\end{array}\right).\end{eqnarray*}

The fourth set of observational data is $12$ Hubble evolution data
from \cite{Simon:2004tf} and \cite{Gaztanaga:2008xz}. Its
$\chi_{H}^{2}$ is defined as
\begin{equation}
\chi_{H}^{2}=\sum_{i=1}^{12}\frac{[H(z_{i})-H_{ob}(z_{i})]^{2}}{\sigma_{i}^{2}}.
\end{equation}
 Note that the redshift of these data falls in the region $z\in(0,1.75)$.

The Big Bang Nucleosynthesis (BBN) data we use here are from
\cite{Serra:2009yp,Burles:2000zk}, whose $\chi^{2}$ is
\begin{equation}
\chi_{bbn}^{2}=\frac{\left(\Omega_{b0}h^{2}-0.022\right)^{2}}{0.002^{2}}.
\end{equation}

And finally for the growth factor data, we define
 \begin{equation}
\chi_{f}^{2}=\sum_{i=1}^{11}\frac{[f(z_{i})-f_{ob}(z_{i})]^{2}}{\sigma_{i}^{2}}.
 \end{equation}
The 11 data of growth factor are summarized in Table
\ref{tab:growth_factor}~\cite{Gupta:2011kw}.
\begin{table}
\begin{centering}
\begin{tabular}{|c|c|c|c|}
\hline $z$  & $f_{obs}$  & $\sigma$  & $Ref$\tabularnewline \hline
$0.15$ & $0.51$  & $0.11$  & \cite{Hawkins} \tabularnewline
\hline$0.22$ & $0.60$  & $0.10$  & \cite{Blake} \tabularnewline
\hline $0.32$ & $0.654$ & $0.18$  & \cite{Reyes} \tabularnewline
\hline $0.35$ & $0.70$  & $0.18$ & \cite{Tegmark} \tabularnewline
\hline $0.41$ & $0.50$  & $0.07$  & \cite{Blake} \tabularnewline
\hline $0.55$ & $0.75$ & $0.18$ & \cite{Ross} \tabularnewline \hline
$0.60$ & $0.73$  & $0.07$ & \cite{Blake} \tabularnewline \hline
$0.77$ & $0.91$  & $0.36$  & \cite{Guzzo} \tabularnewline \hline
$0.78$ & $0.70$ & $0.08$ & \cite{Blake} \tabularnewline \hline $1.4$
& $0.90$  & $0.24$ & \cite{Angela} \tabularnewline \hline $3.0$ &
$1.46$ & $0.29$  & \cite{McDonald} \tabularnewline \hline
\end{tabular}
\par\end{centering}
\caption{\label{tab:growth_factor} Currently available data for
linear growth rate $f_{obs}$ used in our analysis. $z$ is redshift;
$\sigma$ is the $1\sigma$ uncertainty of the growth rate data.}
\end{table}

\subsection{Fitting Results}

\begin{figure}
\begin{centering}
\includegraphics[width=0.8\textwidth]{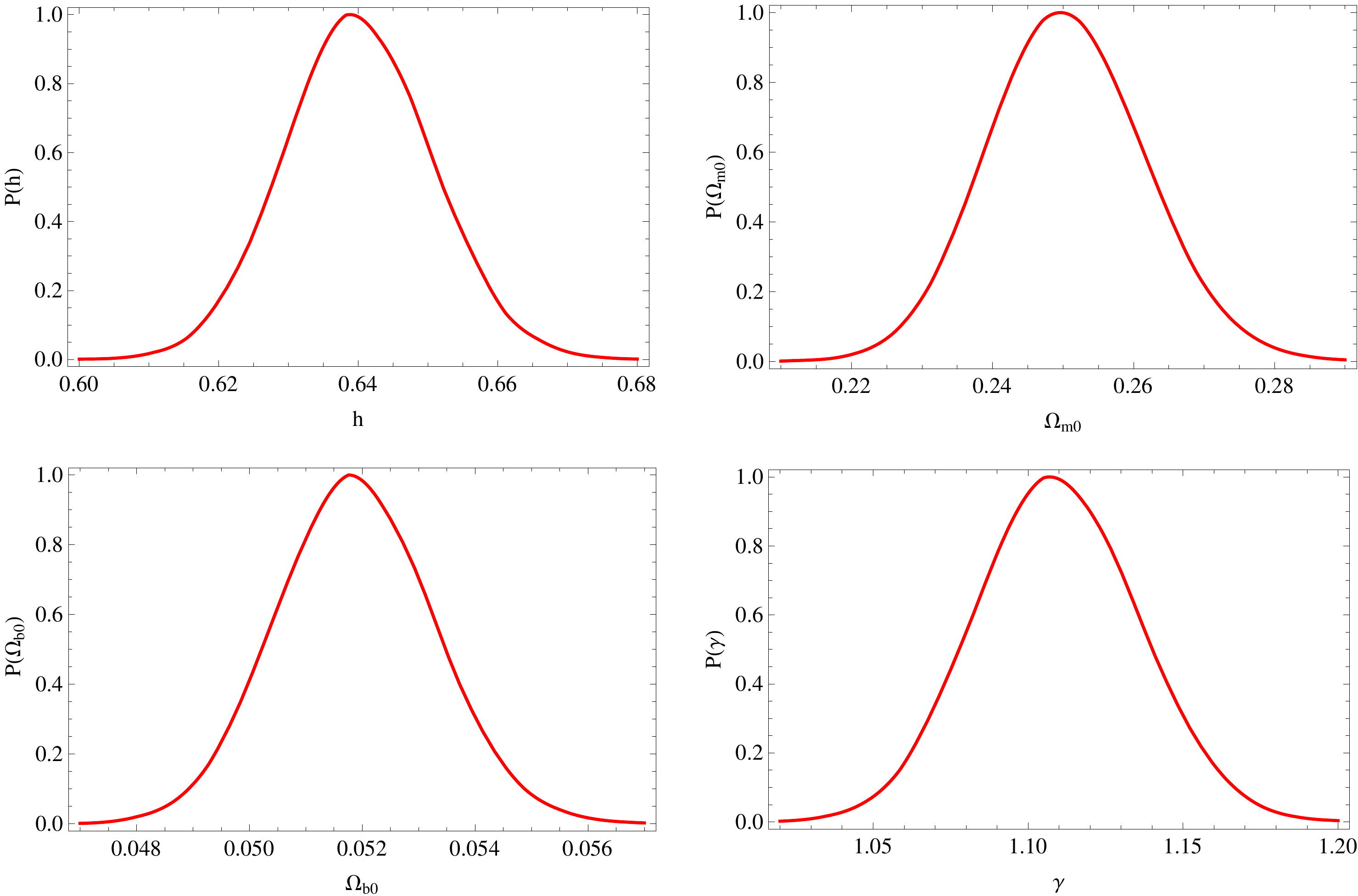}
\par\end{centering}

\caption{\label{fig:likelihood1}1D marginalized distribution
probability of $h,\,\Omega_{m0},\,\Omega_{b0}$ and $\gamma$ using
the full datasets.}

\end{figure}

\begin{figure}
\begin{centering}
\includegraphics[width=0.8\textwidth]{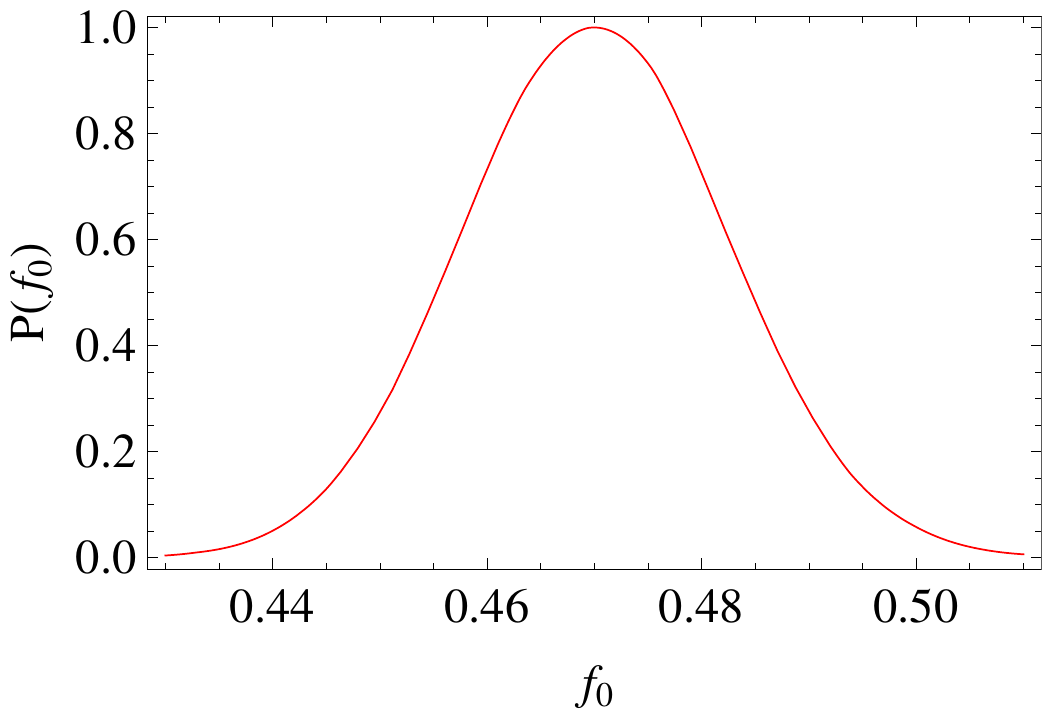}
\par\end{centering}

\caption{\label{fig:likelihoodf} 1D marginalized distribution
probability of $f_0$ using the full datasets.}

\end{figure}

The best fitting values and errors of the model parameters are
summarized in Table~\ref{tab:fit_result}, where we also list the
best fitting values of the corresponding parameters of $\Lambda CDM$
model for comparison. The best fitting  values of $\Omega_{m0}$ and
$h$ are slightly smaller than corresponding ones in the $\Lambda
CDM$ model and the best fitting values of $\Omega_{b0}$ are larger
than corresponding ones in the $\Lambda CDM$ model. We also can see
from Table~\ref{tab:fit_result} that adding the data of growth
factor dose not have much impact on the values of the parameters,
both at $1\sigma$ confidence level and $2\sigma$ confidence level,
which may mean that this model is not sensitive to the linear growth
rate of matter. In addition, we find that $\gamma <1$ is excluded at
$2\sigma$ confidence level, which means that in the ghost dark
energy model, $\beta <0$. Furthermore, we see that the subleading
term $H^2$ of the dark energy density, the early dark energy, could
have a fraction energy density around $10\%$. In Figure
\ref{fig:likelihood1} and Figure \ref{fig:likelihoodf}, we plot the
1D marginalized distribution probability of each parameter using the
full datasets.

\begin{table}
\begin{centering}
\begin{tabular}{|c|c|c|c|}
\hline parameter  & SN+BAO+CMB+H+BBN & SN+BAO+CMB+H+BBN+F  &
$\Lambda CDM$ \tabularnewline \hline \hline $h$ &
$0.642_{+0.012,\,+0.023}^{-0.017,\,-0.025}$  &
$0.642_{+0.010,\,+0.021}^{-0.015,\,-0.027}$  & $0.708$
\tabularnewline \hline \hline $\Omega_{m0}$ &
$0.250_{+0.014,\,+0.026}^{-0.014,\,-0.025}$  &
$0.251_{+0.013,\,+0.026}^{-0.014,\,-0.025}$  & $0.266$
\tabularnewline \hline \hline $\Omega_{b0}$ &
$0.052_{+0.002,\,+0.003}^{-0.002,\,-0.003}$  &
$0.052_{+0.002,\,+0.003}^{-0.002,\,-0.003}$  & $0.045$
\tabularnewline \hline \hline $\gamma$ &
$1.114_{+0.029,\,+0.058}^{-0.035,\,-0.062}$  &
$1.105_{+0.035,\,+0.063}^{-0.028,\,-0.056}$  & $\backslash$
\tabularnewline \hline \hline $f_0$ & $\backslash$ &
$0.473_{+0.012,\,+0.024}^{-0.018,\,-0.029}$  & $0.485$
\tabularnewline \hline
\end{tabular}
\par\end{centering}
\caption{\label{tab:fit_result} The best fitting values within
$1\sigma$ and $2\sigma$ errors for
$h,\,\Omega_{m0},\,\Omega_{b0},\,\gamma$ and $f_0$ for the dark
energy model. The second column shows the results using the datasets
without the data of growth factor, and the third column shows the
results fitted with full datasets. The last column shows the best
fitting results of $\Lambda CDM$ model using the full datasets for
comparison.}
\end{table}

\begin{figure}
\begin{centering}
\includegraphics[width=1\textwidth]{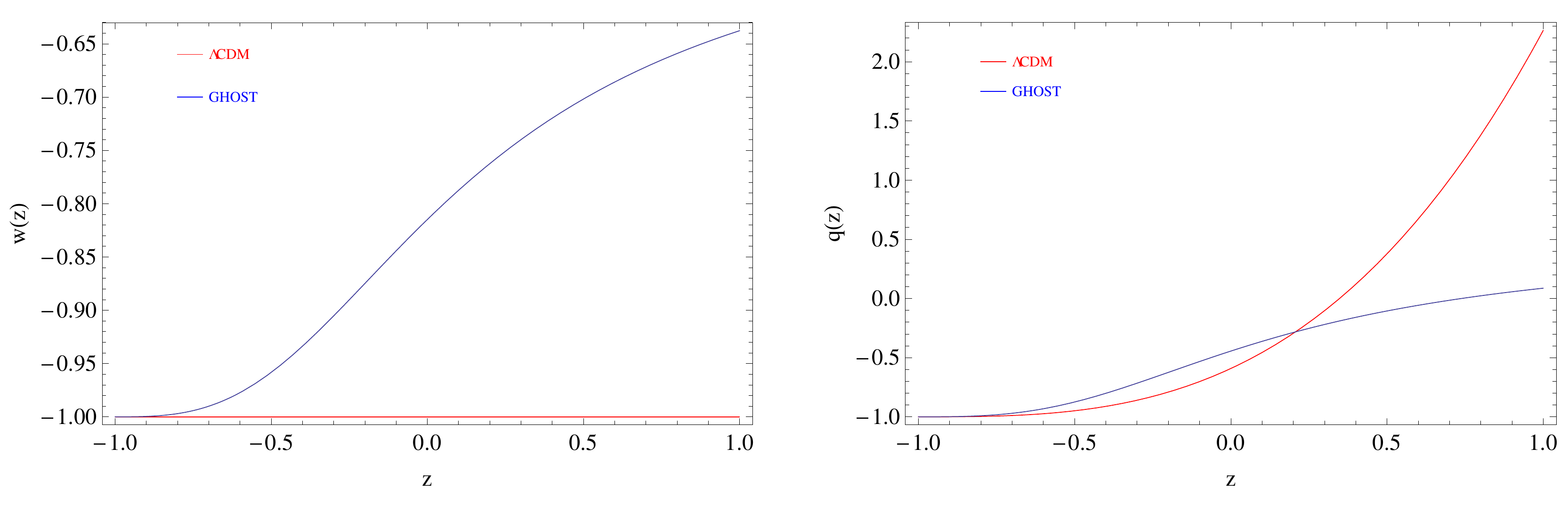}
\par\end{centering}

\caption{\label{fig:evolution}Evolution behaviors of the equation of
state of DE and the deceleration parameter for the QCD ghost dark
energy  model and the $\Lambda CDM$ model.}

\end{figure}

\begin{figure}
\begin{centering}
\includegraphics[width=0.8\textwidth]{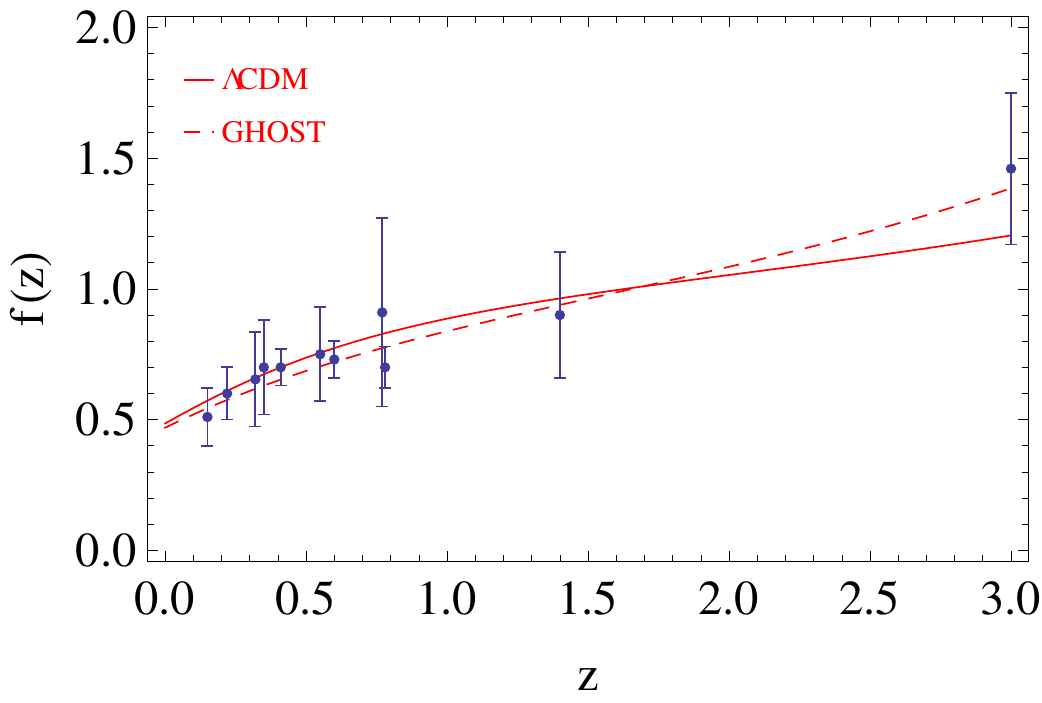}
\par\end{centering}

\caption{\label{fig:gz}Evolution behaviors of the growth factor for
the QCD ghost dark energy model and the $\Lambda CDM$ model.}

\end{figure}

In Figure \ref{fig:evolution} we plot the evolution behaviors of the
equation of state $w(z)$ of DE and the deceleration parameter
$q(z)$, with the best fitting values of the model and the $\Lambda
CDM$ model. In the calculation, we employ the following relations:
\begin{eqnarray}
w(z) & = & -1+\frac{(1+z)}{3H(z)}\frac{dH(z)}{dz},\\
q(z) & = & -1+\frac{(1+z)}{H(z)}\frac{dH(z)}{dz}.\end{eqnarray} The
results show that in the QCD ghost dark energy model, the universe
transits from early matter dominant phase to the de-Sitter phase in
the future, as expected.  The accelerating expansion begins at
$z=0.75$, which is earlier than what the $\Lambda CDM$ model
predicts. $w(z)$ varies from $w>-1$ to $w=-1$ which is similar to
freezing quintessence model~\cite{Caldwell:2005tm}.

The total $\chi^{2}$ of the best fitting values of this model using
the full datasets is $\chi_{min}^{2}=589.422$ for $dof=586$. The
reduced $\chi^{2}$ equals to $1.006$, which is acceptable, but
$\chi_{min}^{2}$ is a little larger than the one for the $\Lambda
CDM$ model, $\chi_{\Lambda CDM}^{2}=558.890$. A similar conclusion
is also reached by other authors using different data set in
\cite{Basilakos:2009wi}. That work studies the dynamics of varying
vacuum energy as a cosmological constant. That is, the equation of
state of the vacuum energy is always kept as $w=-1$. In that case,
there must exist interaction between matter and the vacuum energy.
In Figure \ref{fig:gz}, we plot the evolution of the growth factor
for the QCD ghost dark energy  model and the $\Lambda CDM$ model, it
shows that the ghost dark energy model can not be discriminated by
the data, and that both of these two models fit the data very well,
even the ghost dark energy model looks fitting the data better.


\section{Conclusion and Discussion}

The accelerating expansion (dark energy) of the universe must be
closely related to the vacuum energy of quantum fields. It is
believed that the difference between the vacuum energies in
Minkowski space and in FRW universe might be the origin of observed
dark energy. However, the naive estimate indicates that the
difference should be of the form
$H^2\Lambda_c^2$~\cite{Magg,Maggiore:2011hw}. Such a term is too
small and cannot derive the universe to accelerating expansion. But
this term may play an important role in the early evolution of the
universe, acting as an early dark energy.

On the other hand, the vacuum energy difference from the Veneziano
ghost field introduced in order to solve the so-called $U(1)_A$
problem in QCD has the exact form, $\alpha H+\beta H^2$, where
$\alpha \sim \Lambda_{QCD}^3 \sim (100MeV)^3$. The leading term
gives exactly the order of the observed dark energy. Therefore the
QCD ghost dark energy model is very attractive in the sense that
this model needs not introduce new degrees of freedom or modify
Einstein's general relativity, to explain the accelerating expansion
of the universe observed today.

In this paper, based on the vacuum energy of QCD ghost field,  we
investigated a DE model whose energy density has the form $\alpha
H+\beta H^2$. We studied the dynamical evolution of the QCD ghost
dark energy model and fitted this model with observational data
including SnIa, BAO, CMB, BBN, Hubble parameter and the growth
factor. The best fitting results show that the subleading term of
the energy density makes a negative contribution to the total energy
density.  In this model, the universe transits from early matter
dominant phase to a de-Sitter phase in the future, and the
accelerating expansion begins at $z=0.75$, which is earlier than
that of $\Lambda CDM$ model. The equation of state of DE varies from
$w>-1$ to $w=-1$  like a freezing quintessence model.

The total $\chi^{2}$ of the best fitting values of this model is
$\chi_{min}^{2}=589.422$ for the full datasets with $dof=586$. The
reduced $\chi^{2}$ is $1.006$, which is acceptable, but
$\chi_{min}^{2}$ is a little larger than the one for the $\Lambda
CDM$ model, $\chi_{\Lambda CDM}^{2}=558.890$, for the same datasets.
We further studied the cosmological dynamics of the model by
considering the effect on the growth rate of matter.  The ghost dark
energy model can not be discriminated by the data, and both of this
model and the $\Lambda CDM$ model fit the data very well.

Finally before ending this paper, we would like to stress that in
fact there have not been any precise calculations showing that the
vacuum energy density of the Veneziano ghost of QCD in a FRW
universe is of the form, $\alpha H +\beta^2$,  because the vacuum
energy calculation of the Veneziano ghost is quite difficult in both
flat and curved spacetimes due to the intrinsic difficulties of QCD
and strongly interacting fields in general~\cite{Urban}. But the
vacuum energy calculations of the Kogut-Susskind ghost in 2d QED
(which is the direct analogue of the Veneziano ghost in QCD), in 2d
topological nontrivial spacetime and curved
space~\cite{Urban,Zhitnitsky:2011tr}, and the vacuum energy
calculations of the effective Veneziano ghost of QCD in 4d Rindler
spacetime~\cite{Ohta:2010in} indeed indicate such a power-law
behavior. Note that in those calculations, the kinetic contribution
is  not included, it is expected that the kinetic contribution is of
the same order of magnitude of the potential. While the dynamics of
the effective scalar field model for the Veneziano ghost of QCD in
FRW universe is discussed in the fourth reference in~\cite{Urban},
here we have further made an approximation that the size scale $L$
of some nontrivial manifold is replaced by the Hubble size of the
FRW universe, which becomes time-dependent. In principle, in such an
extension, one has to consider the kinetic contribution of the
time-dependent scale to the vacuum energy density. However, we
assume that such a contribution is subdominant due to slowly
evolution of the scale, compared to the potential term, and absorb
some uncertainties into the two coefficients $\alpha$ and $\beta$,
because at the moment one could not make an explicit calculation to
take into account the effect. Of course, the kinetic contribution of
the Veneziano ghost of QCD and the effect in a time-dependent
spacetime should be seriously studied if the QCD ghost dark energy
model studied in this paper can fit well with the observational
data. Clearly at the moment one just can review this model at a
phenomenological level.


\begin{acknowledgments}
This work is supported in part by grants from NSFC (No. 10821504,
No. 10975168, No. 11035008 and No. 11175077),  by the ministry of
science and technology of China under grant No. 2010CB833004, the
Chinese academy of sciences under grant No.KJCX2-EW-W01 and by the
natural science foundation of Liaoning province with grant No.
20102124.

\end{acknowledgments}


\vspace*{0.2cm}


\end{document}